\documentclass[twocolumn,american,prl,showpacs,superscriptaddress]{revtex4-1}
\usepackage{amsmath}
\usepackage{graphicx}
\usepackage{amssymb}
\usepackage{times}
\usepackage{color}
\usepackage[flushleft]{threeparttable}
\usepackage{pdfpages}
\definecolor{lr}{rgb}{1.0,0.3,0.3}
\definecolor{dg}{rgb}{0.0,0.5,0.0}

\makeatletter

\@ifundefined{textcolor}{}
{%
 \definecolor{BLACK}{gray}{0}
 \definecolor{WHITE}{gray}{1}
 \definecolor{RED}{rgb}{1,0,0}
 \definecolor{GREEN}{rgb}{0,1,0}
 \definecolor{BLUE}{rgb}{0,0,1}
 \definecolor{CYAN}{cmyk}{1,0,0,0}
 \definecolor{MAGENTA}{cmyk}{0,1,0,0}
 \definecolor{YELLOW}{cmyk}{0,0,1,0}
 }

\makeatother

\begin{document}

\title{The role of screening in the density functional applied on transition metal defects in semiconductors}
 

\author{Viktor Iv\'ady}
\email{vikiv@ifm.liu.se} 
\affiliation{Wigner Research Centre for Physics, Hungarian Academy of Sciences,
  PO Box 49, H-1525, Budapest, Hungary}
\affiliation{Department of Physics, Chemistry and Biology, Link\"oping
  University, SE-581 83 Link\"oping, Sweden}

\author{I. A. Abrikosov}
\affiliation{Department of Physics, Chemistry and Biology, Link\"oping
  University, SE-581 83 Link\"oping, Sweden}

\author{E. Janz\'en}
\affiliation{Department of Physics, Chemistry and Biology, Link\"oping
  University, SE-581 83 Link\"oping, Sweden}

\author{A. Gali} 
\email{agali@eik.bme.hu}
\affiliation{Wigner Research Centre for Physics, Hungarian Academy of Sciences,
  PO Box 49, H-1525, Budapest, Hungary}
\affiliation{Department of Atomic Physics, Budapest University of
  Technology and Economics, Budafoki \'ut 8., H-1111 Budapest,
  Hungary}

\pacs{61.72.J-, 61.82.Fk, 71.15.Mb, 76.30.-v}

\begin{abstract}
We study selected transition metal related point defects in silicon
and silicon carbide semiconductors by a range separated hybrid density
functional (HSE06). We find that HSE06 does not fulfill the
generalized Koopmans' Theorem for every defect which is due to the
self-interaction error in the functional in such cases. Restoring the
so-called generalized Koopmans' Condition with a simple correction in
the functional can eliminate this error, and brings the calculated
charge transition levels remarkably close to the experimental data as
well as to the calculated quasi-particle levels from many-body
perturbation theory.
\end{abstract}
\maketitle


Tractable solutions for the problem of many-electron systems are highly
needed in materials science and under intense research. One key
example of such problems is the treatment of small imperfections in
single crystals, i.e.\ point defects in semiconductors. Point defects
can completely change \emph{locally} the chemical bonds of the host
crystal, thus \emph{ab initio} methods are needed in order to
determine their geometry, electronic structure, ionization energies or
optical excitations. The most challenging type of point defects is the
transition metal (TM) related defects in traditional semiconductors
where $d$-orbitals tightly localized on the TM atoms and $sp^3$
hybrid orbitals are both present in the system. It is known from ,
e.g. studies of Mott-insulators that $d$-electron systems may be
highly correlated whereas the electronic structure of traditional
semiconductors can phenomenologically be well-described by independent
particle theories. Simulations of TM defects in semiconductors are
therefore a complex problem, thus it is an obvious system to
investigate the predictive power of the \emph{ab initio} methods.

Density functional theory (DFT) is the most widespread technique for
first principles calculation in condensed matter physics.
Particularly, DFT has been proven to be extremely powerful tool to
study defects in semiconductors \cite{Walle04}.  The success of the
DFT calculations is based on well-developed approximate (semi)local
functionals \cite{C-A80, Perdew81, PW91, PBE} that made it possible to
study relatively large systems at moderate computational cost with a
surprisingly good accuracy \cite{Walle04}. The success of the commonly
used (semi)local functionals might be unexpected as they suffer from
the self-interaction error \cite{Perdew81} which results in the
underestimation of the band gap of semiconductors. The predictive
power of (semi)local functionals is thus restricted for ionization
energies of defects in semiconductors, and can even fail to describe
the nature of the host semiconductor, or the defect state in the
semiconductor correctly in pathological cases \cite{Deak,
  LanyZunger09, Walle09}.  By mixing a non-local Fock exchange into
the density functional, i.e., by using hybrid density functionals, one
can restore the band gap of host semiconductors via tuning the mixing
parameter \cite{Gali03, Knaup05, Gali03PRB68, Gali05, Pasquarello08}.
Alternatively, by introducing a range separated hybrid density
functional (HSE06) it is found that with a fixed mixing parameter and
the range of separation, the band gap of many semiconductors with
$sp^3$ hybrid orbitals can be well reproduced \cite{Kresse06}, and in
addition, the structural parameters, redox reaction energies, and
formation energies of transition metal compounds can also be well
accounted by HSE06 \cite{Ceder10}.  This brings a hope that HSE06 is
able to predict the ground state \cite{Leitsmann} and ionization
energies for TM related defects in semiconductors which contain both
$d$ and $sp^3$ orbitals. However, a consistency of the approach
requires the use of the same mixing and range separation parameters
for the host and impurity states. It has been observed that this
requirement makes it difficult to describe band structure and defect
states simultaneously \cite{LanyZunger10, Walsh08}.

In this \emph{Letter} we study TM defects in silicon (Si) and $4H$
silicon carbide (SiC) semiconductors by HSE06 functional using plane
wave large-supercell calculations, for those experimental data about
their structure and ionization energies are available. We show that
HSE06 can qualitatively fail in such complex systems due to
insufficient screening of the Coulomb interaction between localized
$d$-electrons, leading to incomplete cancelation of the
self-interaction error which manifests as disobeying the generalized
Koopmans' Theorem (gKT) \cite{LanyZunger09, Dabo, Cococcioni,
  LanyZunger10}. We suggest a correction scheme for the {hybrid}
functional which fulfills the conditions of gKT. The corrected
functional allows for a simultaneous description of localized and
extended states, is first-principles in nature, and brings our
theoretical results remarkably close to the experimental data as well
as to the results obtained by means of many-body perturbation
theory. Our work underlines the importance of testing a fulfillment of
gKT for the applied functionals on a given many-electron system.

\begin{figure}
\includegraphics[width=0.95\columnwidth]{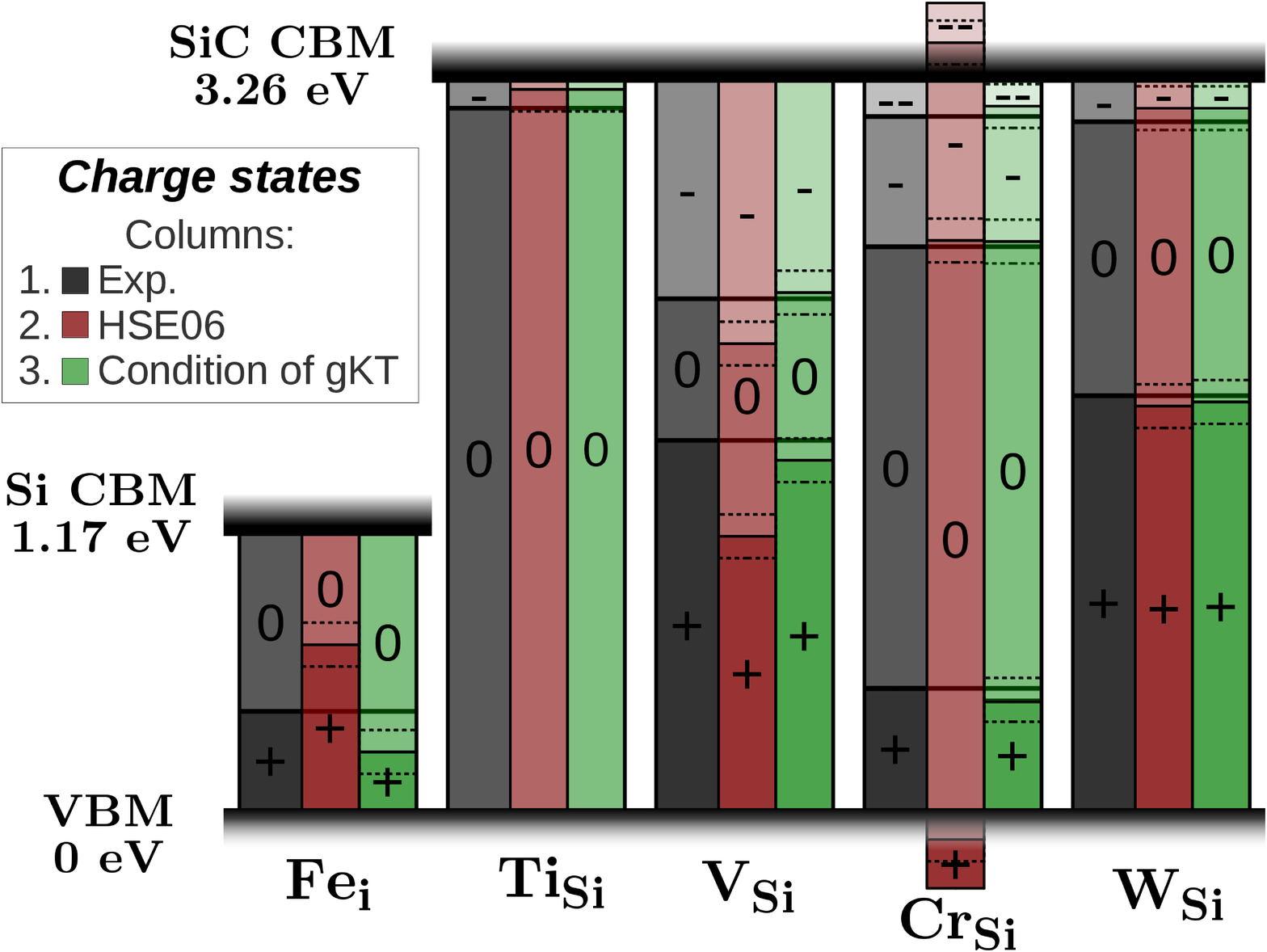}
\caption{\label{fig:fig1}(Color online) Measured and theoretically
  predicted charge transition levels of transition metal impurities in
  $4H$ SiC and Si hosts. In all cases the experimental results
  \cite{Fe_in_Si, SiCsum, Gallstrom09, Beyer} are represented with
  dark gray (first) columns where the thick horizontal line represents
  the charge transition levels. The result of HSE06 calculations and
  our correction method with satisfying gKT are represented with red
  (second) columns and light green (third) columns, respectively,
  where thin horizontal lines provide the calculated charge transition
  levels. The bars indicated by dashed lines around the calculated
  charge transition levels show the inherent uncertainty of the
  calculation. Chromium needs special consideration regarding the
  \mbox{(0\textbar-)} transition level, see text.}
\end{figure}


We selected interstitial Fe in Si (Fe$_\text{i}$), and titanium
(Ti$_\text{Si}$), vanadium (V$_\text{Si}$) and chromium
(Cr$_\text{Si}$) substituting Si-site in $4H$ SiC. For all of these
defects, the structure and the corresponding ionization energies are
well established from experiments
\cite{Fe_in_Si,SiCsum,Gallstrom09,Lambrecht06}. In addition, we
investigate tungsten (W) in $4H$ SiC because W-related ionization
energies have been detected recently \cite{Beyer}. In this case, the
origin of the W-related centers are not ambiguously
identified. According to our previous study \cite{Ivady}, both
Si-substitution site (W$_\text{Si}$) and asymmetric split vacancy
configuration (W$_\text{ASV}$) may be stable, thus both defects are
considered.  For the electronic structure calculations we use
\textsc{VASP} package \cite{VASP} with plane wave basis (cut-off:
420~eV) and projector augmented-wave (PAW) \cite{PAW} method for the
ion-cores. In the case of metal atoms we utilize small core PAW
projectors. In order to model $4H$ SiC and Si host we use a 576-atom
and 512-atom supercells, respectively. In the large supercells the
Brillouin zone is sampled at $\Gamma$-point. In the case of charged
defects the size dependence of total energies is eliminated by charge
correction \cite{Freysoldt} with the order of magnitude 0.1~eV in our
supercells. The geometry of the defects is optimized unless the forces
acting on the atoms are less than 0.01~eV/\AA .

In order to examine the performance of the HSE06 screened non-local
hybrid density functional on the selected transition metal defects in
semiconductors
we calculate the adiabatic
ionization energies or charge transition levels of these defects,
\begin{equation}
 \varepsilon \left ( q | {q}' \right ) = \left ( E_{{q}'} +
 E^\text{corr} \left ( {q}' \right ) \right ) - \left ( E_{q} +
 E^\text{corr}\left ( q \right ) \right ) - \varepsilon_\text{CBM}
\end{equation}
where $E_{q}$ is the total energy and $E^\text{corr}(q )$ is the
charge correction of charge state $q$ and $\varepsilon_\text{CBM}$ is
the conduction band minimum (see Ref.~\onlinecite{Walle04} for more details).
The HSE06 functional gives the proper
description of the band structure of the semiconductor host. The band
gap error is within 0.1~eV for the pristine $4H$ SiC and Si host which
causes uncertainty in the calculated $\varepsilon \left ( q|{q}'
\right )$ of about 0.1~eV. By comparing the experimental data with the
results of HSE06 calculations it is apparent from Fig.~\ref{fig:fig1}
and Table~\ref{tab:corr} that HSE06 cannot predict the correct charge
transition levels for all the defects. The magnitude of the
discrepancies depend on the TM as well as the charge transition level
(Table~\ref{tab:corr}). Particularly, the calculated ionization
energies of Cr strikingly differs from the experimental values, and
are \emph{qualitatively} wrong.

\begin{table}
\begin{threeparttable}
\begin{ruledtabular}
\caption{\label{tab:corr} (Color online) Experimental values of charge
  transition levels ($\varepsilon_{\textrm{exp}}$), deviations of the
  theoretically predicted values from the experiment
  for transitional metal related point
  defects, Ti$_{\textrm{Si}}$, V$_{\textrm{Si}}$, Cr$_{\textrm{Si}}$
  and W$_{\textrm{Si}}$ in $4H$ SiC and Fe$_{\textrm{i}}$ in Si. In
  the calculations HSE06 functional and a corrected HSE06 functional
  (see text) are used. The finite values of 
  $E_{\text{NK}}$ clearly indicates the presence of self-interaction
  error in the HSE06 functional ($\Delta \varepsilon_{\text{HSE06}}$).
  When small value of $E_{\text{NK}}$ is obtained (${E}'_{\text{NK}}$)
  with the help of additional correction functional (V$_\text{w}$) with a
  strength parameter $w$, the discrepancy ($\Delta
  \varepsilon_{\text{HSE06+V}_\text{w}}$) goes below 0.1~eV which is the
  uncertainty in our calculation. Light red (dark green) numbers
  represent the case when the calculated charge transition level is
  out of (within) the error bar. 
  The unit of the data is eV in all columns. }
 \begin{tabular}{ l|c|c|c|ccc}
  Transition            & $\varepsilon_{\textrm{exp}}$ &
  $E_{\text{NK}}$  & $\Delta \varepsilon_{\text{HSE06}}$  & $w$ & ${E}'_{\text{NK}}$ & $\Delta \varepsilon_{\text{HSE06+V}_\text{w}}$ \\
  \hline
  Ti$_{\textrm{Si}}$ : (0\textbar -)   & -0.12$^a$ &
  \textcolor{dg}{-0.02} &  \textcolor{dg}{+0.01} &  0.0 & \textcolor{dg}{-0.02}  & \textcolor{dg}{+0.10}\\
  V$_{\textrm{Si}}$ : (+\textbar 0)    & -1.60$^a$ & \textcolor{lr}{-0.91} & \textcolor{lr}{-0.43} & -2.7 & \textcolor{dg}{+0.02}  & \textcolor{dg}{-0.09}\\
  V$_{\textrm{Si}}$ : (0\textbar -)    & -0.97$^a$ & \textcolor{lr}{-0.81} & \textcolor{lr}{-0.20} & -2.2 & \textcolor{dg}{+0.02}  & \textcolor{dg}{+0.03}\\
  Cr$_{\textrm{Si}}$ : (+\textbar 0)   & -2.70$^a$ &   -- &  VB  &  -3.0 & \textcolor{dg}{-0.00}  & \textcolor{dg}{-0.05}\\
  Cr$_{\textrm{Si}}$ : (0\textbar -)   & -0.74$^a$ & \textcolor{dg}{-0.21} & \textcolor{lr}{+0.02} &  --  &  --  & \textcolor{dg}{+0.01}\\
  Cr$_{\textrm{Si}}$ : (-\textbar 2-)  & -0.18$^a$ & \textcolor{lr}{-1.88} & \textcolor{lr}{+1.85} & -6.0 & \textcolor{dg}{+0.01}  & \textcolor{dg}{+0.05}\\
  W$_{\textrm{Si}}$ : (+\textbar 0)    & -1.40$^b$ & \textcolor{dg}{-0.13} & \textcolor{lr}{-0.05} & -1.2 & \textcolor{dg}{+0.01}  & \textcolor{dg}{-0.03}\\
  W$_{\textrm{Si}}$ : (0\textbar -)    & -0.18$^b$ & \textcolor{dg}{-0.03} & \textcolor{dg}{+0.06} &  0.0 & \textcolor{dg}{-0.03}  & \textcolor{dg}{+0.06}\\
  Fe$_{\textrm{i}}$ : (+\textbar 0)    & -0.79$^c$ & \textcolor{lr}{-0.68} & \textcolor{lr}{+0.32} & -3.8 & \textcolor{lr}{-0.09}  & \textcolor{lr}{-0.15}\\
 \end{tabular}
\end{ruledtabular}
 \begin{tablenotes}
  \item $^a$ Ref. \cite{SiCsum}, $^b$ Ref. \cite{Beyer}, $^c$ Ref. \cite{Fe_in_Si}
 \end{tablenotes}
\end{threeparttable}
\end{table}


Let us demonstrate that the failure of HSE06 for some defects
originates from the incorrect treatment of atomic-like $d$-orbitals of
TM atoms. It is well-known that self-Hartree and the exchange
potentials do not cancel each in (semi)local DFT functionals.  In
(semi)local functionals the resultant self-repulsive potential lower
the localization degree of the states and cause the spurious
occupation dependence of the Kohn-Sham energies as well as the
spurious, generally convex, curvature of the total energy with respect
to fractional occupations \cite{LanyZunger09,Dabo,Cococcioni}.  The
opposite behavior is characteristic for Hartree-Fock (HF) method.  By
mixing HF and semi-local exchange, the self interaction error is
reduced in hybrid functionals. Unfortunately, it is not eliminated
completely.

Previous studies show
\cite{Dabo,Cococcioni,LanyZunger09,LanyZunger10} that one of the
quantitative manifestation of the self-interaction error in DFT
functionals is the discrepancy between the Kohn-Sham eigenvalue of the
highest occupied state and the corresponding ionization energy. This
energy difference is usually called as Non-Koopmans' energy,
\begin{equation}
 E_\text{NK} = \varepsilon_{N} - E_{I} = \varepsilon_{N} - \left ( E_{N} - E_{N-1} \right)
\end{equation} 
where $\varepsilon_{N}$ is the Kohn-Sham energy of a localized state
in $N$ electrons system, $E_{I}$ is the ionization energy of the
system which equal to the difference of the total energies of the $N$
electron, $E_{N}$, and $N-1$ electron system, $E_{N-1}$. In HF theory
the Koopmans' Theorem states that the single particle energies equal to
the ionization energy in case of every occupied states, while in DFT
this statement is only valid for the highest occupied state with an
exact exchange-correlation functional. In the case of exact functional
$E_\text{NK} = 0$ condition should be fulfilled, the so-called
generalized Koopmans' Condition (gKC). Due to the Janak's and Slater's
theorems, other important features of the exact functional is
simultaneously remedied: linear behavior of the total energy and
constant behavior of (highest occupied) single particle level with
respect to fractional occupation are also fulfilled. The last
condition explicitly shows that the single particle state does not
suffer from any occupational dependent potential, thus represents a
self-interaction free functional.

In Table~\ref{tab:corr} one can see the difference in the calculated
HSE06 results and experimental data as well as the calculated
Non-Koopmans' energy for the defect state that gets occupied through
the charge transition. In the calculation of $E_\text{NK} $ we
considered the two charge states of the defect with fixing the
geometry as found in the first charge state. The results clearly show
that this quantity correlates with the discrepancy between HSE06
results and experimental data for the charge transition levels. The
obvious conclusion of this finding is that the errors are originated
from the spurious self-interaction which still remains in the HSE06
functional.

In HSE06 functional, the mixing parameter and the
range of separation play the role of effective screening of Coulomb
interaction \cite{Marques11, Scanlon11, Iori12}, and their variation
in practice interpolates between the substantial underestimation of
the localization effects in (semi)local functionals and their
overestimation in Hartree-Fock theory. But the strength of the
screening should itself depend on the localization degree of the
states. In the case of TM impurities the strong repulsive interaction
between the electrons stems the extension of the atomic-like electron
states. In such a strongly correlated system, the localized states are
more favorable than the extended one.  Due to the large localization
degree of the $d$-states with respect to that of host $sp^3$ hybrid
orbitals, in general it should not be possible to describe both
subsystems with the same values of parameters in a hybrid functional.
Lany and Zunger \cite{LanyZunger10} demonstrated that one should not
expect that adjusting parameters of hybrid functionals would
automatically give accurate results for both, host band gaps and
defect levels, and pointed out homogeneous screening of the Fock
exchange in hybrid functionals as serious simplification. Still, it is
desirable to have an option to describe the orbitals with different
degrees of localization in calculations.

At this stage, it is important to mention that, a systematic study of
the role of nonlocal exchange in the electronic structure of
correlated oxides by F. Iori \emph{et al.} \cite{Iori12} demonstrates
that the default values of parameters of HSE06 functional
systematically {\it overestimate} band gaps in these materials,
indicating insufficient screening of Coulomb interaction. Negative
values of $E_\text{NK}$ in Table~\ref{tab:corr} also indicate that the
screening of Coulomb interaction is insufficient \cite{LanyZunger10}.

In order to prove the self-interaction error of HSE06 functional, we
carry out sophisticated parameter free GW$_{0}$ calculations
\cite{Hedin} on defective supercells with 128 atoms and HSE06
relaxed geometry. We apply 1344 bands in the calculation of the
response function, and the Brillouin zone is sampled with
$2\times2\times2$ Monkhorst-Pack $k$-point set \cite{MP76}. The
starting wave functions are obtained from HSE06 calculation, then the
Green-function $G$ and the wave functions are self-consistently
updated while keeping the screened Coulumb-interaction $W$ fixed. We
find that four iterations are sufficient to reach the self-consistent
quasi-particle levels within 0.05~eV. The observed quasiparticle
correction (see Table~\ref{tab:singlev} ) confirmed the
over-localization error of the HSE06.
\begin{table}
\begin{ruledtabular}
\caption{\label{tab:singlev} The position of the highest occupied
  single particle levels of transitional metal defects in the gap with
  respect to the conduction band edge (CBM) in eV unit. The table
  shows the results of calculations with  GW$_0$ method, HSE06
  functional and corrected HSE06 functional (HSE06+V$_\text{w}$).  
  In the case of
  Cr$_{\textrm{Si}}$(0) the corresponding single particle state falls
  in the valence band shown as ``VB'', thus GW$_0$ started far from
  the true ground state of the system.  When Non-Koopmans' energy
  ($E_{\text{NK}}$) is not negligible the correction method as well as
  the GW$_0$ method shift the levels up with approximately the same
  value. This result confirms the presence of self-interaction error
  in the HSE06 functional and validate our corrected functional.}
 \begin{tabular}{ l|cc|cc }
  Defect 		& GW$_0$ & HSE06 & HSE06+V$_\text{w}$ \\
  \hline
  V$_{\textrm{Si}}$(0)	& -1.86  &  -2.79 & -1.94 \\
  V$_{\textrm{Si}}$(-)	& -1.21  &  -2.19 & -1.50 \\
  W$_{\textrm{Si}}$(0)	& -1.66  &  -1.85 & -1.68 \\
  W$_{\textrm{Si}}$(-)	& -0.94  &  -0.83 & -0.83 \\
  Cr$_{\textrm{Si}}$(0)	& -3.22  &   VB   & -2.83 \\
  Cr$_{\textrm{Si}}$(-)	& -0.74  &  -1.00 &   --  \\
  Cr$_{\textrm{Si}}$(2-)& +0.07  &  -0.49 & -0.17 \\
 \end{tabular}
\end{ruledtabular}
\end{table}

Considering $E_\text{NK}$ is an appropriate measure to construct a
self-consistent correction method to eliminate the self-interaction
error in the functional \cite{LanyZunger09,Dabo}, we propose a scheme
that does not use any empirical parameters to counteract the error due
to insufficient screening of the Coulomb interaction between
$d$-electrons localized at TM related defects, which is apparently
present in HSE06 functional.

In our correction technique the occupation dependent potential
\begin{equation}
\label{eq:Vw}
 V_w^{Im} = \frac{w}{2} \left ( 1-2n_m^I \right )
\end{equation}
is applied on the $d$-orbitals together with the screened non-local
HSE06 functional. In Eq.~(\ref{eq:Vw}) $w$ is a parameter
for the strength of the potential, which physical meaning will become
clear below, and which is determined self-consistently by satisfying
gKC. We emphasize that the resultant functional is \emph{ab initio} in
the sense that it is not fit to any empirical parameter. We note that
the form of Eq.~(\ref{eq:Vw}) is equivalent to Dudarev's
implementation of LDA+U method \cite{Dudarev}, and therefore it is
straight-forward to use the suggested scheme with existing
first-principles package like VASP.
The occupation number $n_m^I $ in \textsc{VASP} implementation
is determined as projection of the wave function of the system on the
spherical harmonics $lm$ on site of atom $I$, where $l=2$ in our case.

The results are summarized in Table~\ref{tab:corr}.  One sees
that gKC could be fulfilled with great accuracy (${E}'_{\text{NK}}$)
in all cases except for Fe$_\text{i}$ in Si and Cr$_{\textrm{Si}}$ :
(0\textbar -) in SiC. In case of iron $E_\text{NK}$ has an extrema and
does not reach zero. We observed that the particular defect state is
not so well localized on the Fe $d$-orbital in Fe$_\text{i}$ defect,
thus our correction scheme is not sufficiently effective. Similar
effect can be found in case of negatively charged Cr$_{\textrm{Si}}$
where in the ground state three electron occupy the $e$ and $a_1$
levels with parallel spins. The highest occupied $a_1$ level is also
less atomic-like compared to the $e$ level. In Table~\ref{tab:corr}
the Cr$_{\textrm{Si}}$:(0\textbar -) charge transition level is
calculated as $\varepsilon_{GW} + E^\text{relax} +E^\text{corr}$,
where $\varepsilon_{GW}$ is the highest occupied quasi-particle level
in the negative charge state, and $E^\text{relax}$ is the relaxation
energy of the defect due to ionization \cite{Suppl}. Note that achieving gKC
with a certain accuracy brings the corrected single particle level to
be independent from the occupation number with the same accuracy
indicating the self interaction free description of the defect
state. The charge transition level was obtained after the geometry
relaxation of system with the use of the corrected HSE06+V$_\text{w}$
functional shown in the last column of
Table~\ref{tab:corr}. Apparently, this correction method can bring the
theoretical values closer to the experimental values in every
case. When gKC is accurately satisfied the calculated levels approach
the experimental levels within 0.1~eV which is the expected
uncertainty in our calculation. 
It also can be seen that when gKC is fulfilled the position of
the defect state in the gap is close to the corresponding
quasi-particle level calculated by the GW$_0$ approximation of the
many-body perturbation theory. We note here that the formation
energies of the defects are also affected by our correction scheme
which is discussed in the Supplemental Material \cite{Suppl}.

\begin{figure}
\includegraphics[width=0.95\columnwidth]{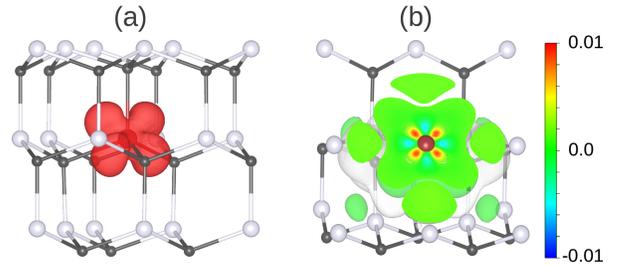}
\caption{\label{fig:fig2}(Color online) (a) The calculated electron
  density (with an isosurface value of 0.05) of the unpaired
  $d$-orbital of the neutral vanadium substitutional defect in $4H$
  SiC. (b) 2D plot of the differences between two calculated charge
  densities using HSE06 functional and the corrected HSE06 functional
  around the vanadium atom. The positive finite density clearly
  represents the overlocalization error of the HSE06 functional.}
\end{figure}

Considering the strength of the required correction potentials an
important remark can be made.  In all the cases when the correction is
needed, the parameter $w$ is negative representing a repulsive
potential for occupied states. This observation confirms that HSE06
functional over-localizes the defect state of the TM impurities, while
this spurious effect is eliminated with our technique.  In more
details this is illustrated in Fig.~\ref{fig:fig2}(b) where the
difference of the spin density of HSE06 and HSE06+V$_\text{w}$ calculation is
shown.  In the case of TM impurities, the HSE06 functional clearly
represents an over-correction with respect to standard (semi)local DFT
functionals.

In order to understand the physical meaning of parameter $w$ in
Eq.~(\ref{eq:Vw}), we will make use of the similarity between the way
how hybrid functionals and LDA+U approach \cite{Anisimov91,Dudarev,
  Liechtenstein} correct the self-interaction error, pointed out by
several groups \cite{Iori12}.  For the case of correlated electrons it
becomes particularly apparent within the so-called exact exchange for
correlated electrons method by Novak and co-workers \cite{Novak06}.
LDA+U functional was introduced in order to take into account the
effects of the strong correlation which is usually not properly
described by (semi)local functionals
\cite{Anisimov91,Dudarev,Liechtenstein}.

The positive value of the Hubbard parameter $U$ representing the
strength of the \emph{screened} Coulomb potential makes the atomic
like states more favorable and increase the localization of the states
on atomic orbitals. In practice the screening of $U$ is achieved by a
reduction of the large bare Coulomb value, and the actual strength of
the on-site electron repulsion can be either determined empirically
\cite{Andersson07,Alling10}, or evaluated \emph{ab initio}
\cite{Aryasetiawan04,Miyake08}.  As a matter of fact, in our case of
the TM defects the Coulomb interaction between localized electrons
(Fig.~\ref{fig:fig2}) appears to be insufficiently screened in HSE06
functional due to its homogeneous nature and parameters adjusted to
describe more extended states. In LDA+U formalism this error could be
corrected by a reduction of $U$ parameter. Employing the above
mentioned similarity between the hybrid functionals and LDA+U method,
we see that the action of occupation dependent potential in
Eq.~(\ref{eq:Vw}) on localized orbitals at impurity sites effectively
increases the screening of Coulomb interaction between electrons
occupying these orbitals.

Finally, we show the predictive power of our methodology demonstrated
on W-related defects in $4H$ SiC. We find that W$_{\textrm{Si}}$
defect has two charge transition levels and these are very close to
the experimental values. In the case of W$_\text{ASV}$ defect HSE06
again suffers from the self-interaction error, $E_\text{NK}$=-0.18~eV,
and correction is needed in the functional with $w$=-2.4~eV. It is
worthy to notice that this value differs from the value needed for
W$_{\textrm{Si}}$ defect which shows that the self-interaction error
for TM defects may be different in various defect configurations. With
our correction method the \mbox{(0\textbar-)} charge transition level of
W$_\text{ASV}$ defect is at $\varepsilon _\text{CBM}$-0.64~eV where no
W-related charge transition level was found in the experiments (see
Table~\ref{tab:corr}). This indicates that W$_{\textrm{Si}}$ is
associated with the measured W-related center \cite{Beyer}, thus we
show the calculated ionization energies of W$_{\textrm{Si}}$ in
Table~\ref{tab:corr}.


In summary, our theoretical investigation revealed that the HSE06
hybrid functional may overlocalize the defect states arisen from
transition metal defects. The reason of this error is the
self-interaction that is increased due to the insufficient screening
of Coulomb interaction between electrons localized at the transitional
metal defects. By invoking the generalized Koopmans' Condition we were
able to indicate the presence of self-interaction error in the
calculation. Furthermore, we suggest a correction method which can
eliminate the self-interaction error of hybrid functionals in the case
of atomic-like defect states. With this technique all the calculated
charge transition levels become accurate within 0.1~eV. In addition,
the corrected Kohn-Sham levels are close to those obtained by the
sophisticated but computationally demanding GW$_0$-method.


Discussion with Rickard Armiento are highly appreciated. Support from
the Swedish Foundation for Strategic Research, the Swedish Research
Council, the Swedish Energy Agency, the Swedish National
Infrastructure for Computing Grants No. SNIC 011/04-8 and
No. SNIC001-10-223, Knut \& Alice Wallenberg Fund, EU FP7 project
DIAMANT, and the European Science Foundation for Advanced Concepts in
ab-initio Simulations of Materials is acknowledged.


%

\newpage


\section*{Supplementary material}

states of this TM-atom, or in other words, on the charge state of the defect. Due to the varying additional screening functional the total energy will be not "unique". The main difficulty may arise from this method when the total energies of different configurations of the given TM impurity are compared in order to determine their relative stability.

To overcome this difficulty, it is possible to choose a common screening parameter $w_c$ for two or more neutrally charged defects. A rational choice is the average of the parameters $w$ obtained for different configurations of the given TM impurity, however it is unknown how sensitive is the relative stability to the choice of the common parameter $w_c$.

We studied the formation energies of the experimentally most relevant TM defects in 4H-SiC. We considered the substitutional defect (TM$_{Si}$) at $h$ site and the asymmetric split vacancy defect (ASV), consisting of TM$_{Si}$ and a carbon vacancy (TM$_{Si}$-V$_C$) at $hh$ site when TM is V or W.

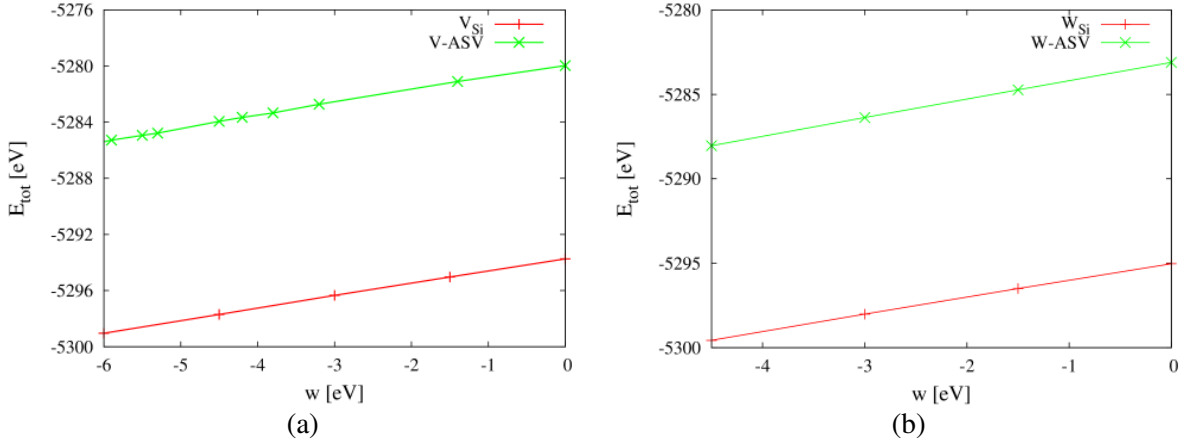

FIG. S1. (Color online) The total energy ($E_{tot}$) of substitutional and ASV defects of neutral (a) vanadium (V) and (b) tungsten (W) is plotted as a function of the applied screening potential ($w$).

The total energies of the two types of TM-defect with respect to screening parameter $w$ can be seen in Figs. S1(a) and (b). Total energies depend linearly on the parameter $w$ and, in addition, the slope of the curves is close to each other, for example 0.912 for V-ASV and 0.885 for V$_{Si}$. Due to this behavior the relative stability is approximately (within ~0.1 eV) independent on the choice of the common screening parameter $w_c$ and is close to results of HSE06 calculation. In critical cases (e.g. when the difference in total energies is close to 0.1 eV), the hierarchy may change as a function of $w$ parameter. W impurity is such an example in *4H* SiC.

The obtained screening parameter and the corrected charge transition levels for the considered defects are summarized in Table S1.

TABLE SI. The optimized screening parameter $w$ that fulfill the gKC and the corresponding corrected charge transition levels with respect to the conduction band edge ($\Delta E_{HSE06+Vw}$) of substitutional and ASV defects of V and W.

| defect | $w$ [eV] | $\Delta E_{HSE06+Vw}$ [eV] |
|---|---|---|
| $V_{Si}$ : (+\|0) | – 2.7 | – 1.69 |
| $V_{Si}$: (0\|-) | – 2.2 | – 0.94 |
| V-ASV : (+\|0) | – 4.6 | – 1.47 |
| V-ASV : (0\|-) | – 4.2 | – 1.29 |
| $W_{Si}$ : (+\|0) | – 1.2 | – 1.43 |
| $W_{Si}$: (0\|-) | 0.0 | – 0.12 |
| W-ASV : (+\|0) | – 2.6 | –2.44 |
| W-ASV : (0\|-) | – 2.4 | – 0.64 |

The chosen common parameters and the corresponding relative stability at Si-rich condition can be found in Table SII. With this data it is possible to draw the transitional level diagram which shows the relative stability with respect to the Fermi-level (Figure S2). We note that the HSE06+$V_w$ functional with the use of common $w_c$ parameter produces superior results in comparison to conventional calculations using HSE06 functional.

TABLE SII. The chosen common screening parameters $w_c$ for the pair of substitutional and ASV defects of V and W. The relative stability ($E_{rel}$) of neutral defects is shown as obtained by HSE06 and HSE06+$V_w$ functionals.

| defect pairs | $w_c$ [eV] | $E_{rel,HSE06}$ [eV] | $E_{rel,HSE06+V_w}$ [eV] |
|---|---|---|---|
| $V_{Si}$ – V-ASV | – 3.65 | 1.82 | 1.87 |
| $W_{Si}$ – W-ASV | – 1.9 | 0.03 | – 0.15 |

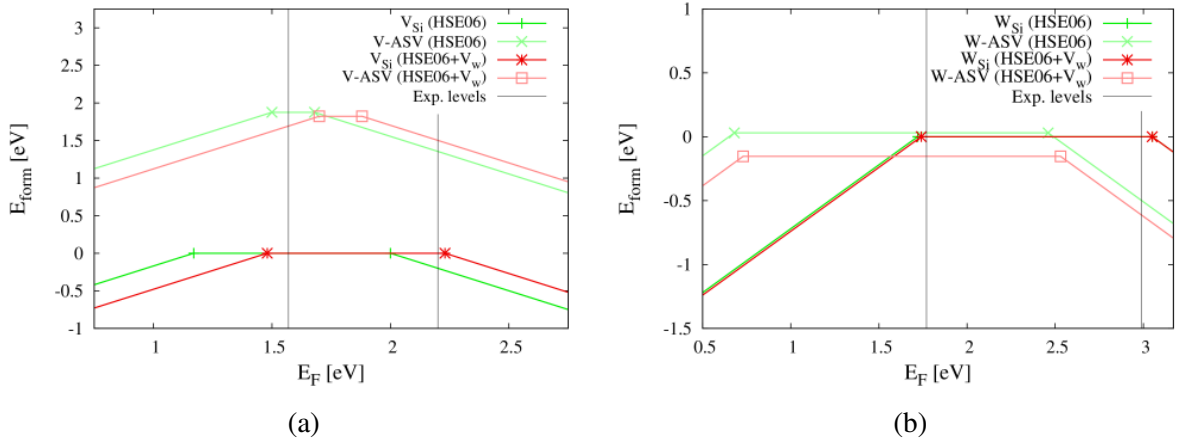

FIG. S2. (Color online) Charge transition level diagram of two different types of defect of vanadium (V) (a) and tungsten (W) (b) at Si-rich condition as obtained by HSE06 and HSE06+$V_w$ functionals (see text). Black vertical lines represent the position of the experimental charge transition levels. The formation energy of the substitutional defect was set to zero as a reference energy in both cases.

## 3. Partial density of states

Partial density of states of neutrally charged substitutional vanadium in 4H-SiC is plotted in Fig. S3. Calculations are carried out using both HSE06 and HSE06+$V_w$ functionals. In the case of vanadium ($V^{4+}$) there is only one electron to occupy the lowest double degenerate $d$-like state ($e$) that splits into two non-degenerate levels in $C_{1h}$ symmetry. The major spin channel is chosen to be the spin-up channel. In the major spin channel both states appear in the gap, the lowest is occupied and the other one is unoccupied as the Fermi-level is at 0.37 eV and 1.27 eV in HSE06 and HSE06+$V_w$ calculations, respectively. The states are slightly hybridized with the $s$ and $p$ orbitals of the host semiconductor. The other $d$-states, however, are so strongly hybridized with the $s$ and $p$ orbitals that they lose their atomic character. This is particularly apparent for the occupied states below the Fermi-level [see Fig. S3(c)] while empty resonant $d$-states may occur in the conduction band. The effect of the correction potential on those states is almost negligible but very strong on the states occurring in the band gap.

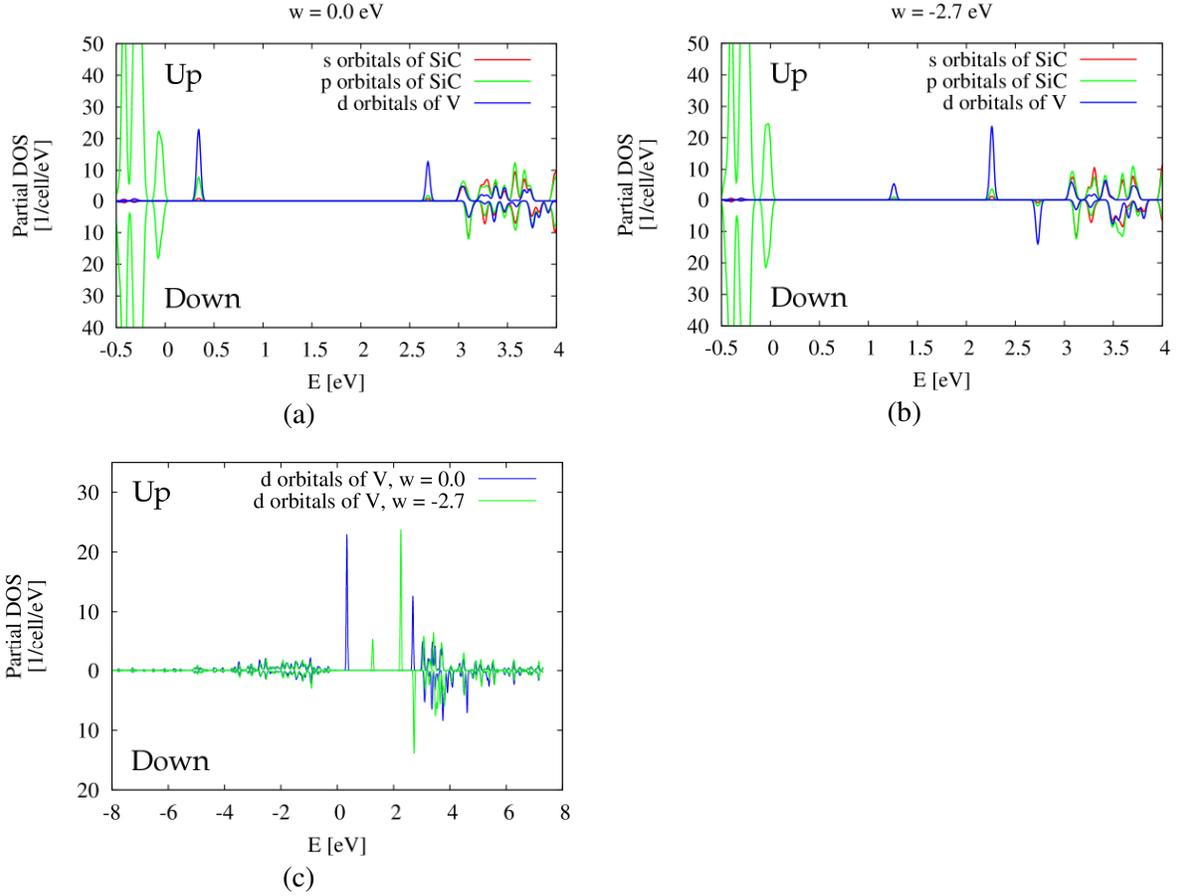

FIG. S3. Partial density of states substitutional vanadium in 4H-SiC calculated by (a) HSE06 functional and (b) and HSE06+$V_w$ functional close to the Fermi-level; (c) by HSE06 and HSE06+$V_w$ functional in wide energy range. The energy of the valence band edge was chosen to zero as reference. Up(down) labels the spin-up(down) channel. The spin-up channel is the majority spin channel. The calculated Fermi-level is at 0.37 eV and 1.27 eV in HSE06 and HSE06+$V_w$ calculations, respectively.

## 4. Calculation of the electronic structure of chromium

We mention here that chromium (Cr) defect needs special consideration in determining $w$. In the ground state of the negatively charged $Cr_{Si}$ defect a double degenerate $e$ level and an $a_1$ level appear in the gap, originated from the split $d$-orbitals of Cr in the crystal field of *4H*-SiC with $C_{3v}$ symmetry. In the neutral charge state two electrons occupy the $e$ state while the $a_1$ state is empty, forming $S = 1$ spin state. Due to the strong correlation of the $d$-orbitals $S = 3/2$ state is formed in the negatively charged state where three electrons occupy the $e$ and $a_1$ levels with parallel spins. In the negatively charged state, our correction method acts on the highest occupied $a_1$ level which is less atomic like compared to the $e$ level. Our correction method makes the atomic-like defect states more or less favorable therefore

5

changes the charge density in a way that changes the character of the $a_1$ level. In Table I of the main text the Cr$_{Si}$:(0|–) charge transition level is calculated as $\varepsilon_{GW} + E^{relax} + E^{corr}$, where $\varepsilon_{GW}$ is the highest occupied quasi-particle level in the negative charge state, and $E^{relax}$ is the relaxation energy of the defect due to ionization. In the case of the Cr$_{Si}$:(–|2–) transition, the $a_1$ level becomes empty and the $e$ level gets fully occupied by ionizing from (–) to (2–) states. Here, we set $S = 1/2$ instead of the energetically favorable $S = 3/2$, for the initial negative charge state in order to be able to study of the effect of the occupation of $e$ level and to correct its level *via* satisfying gKC with our correction functional.

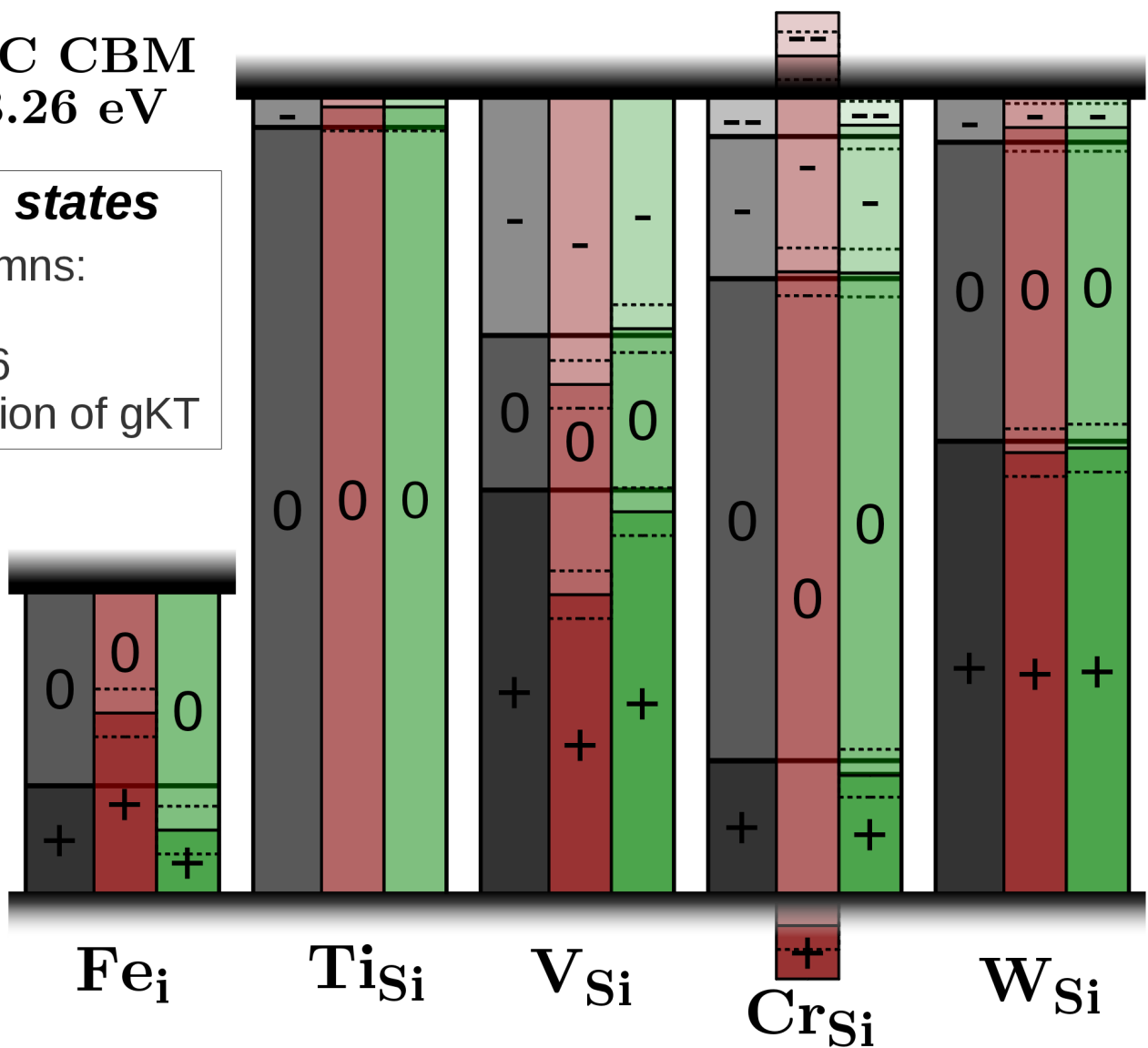

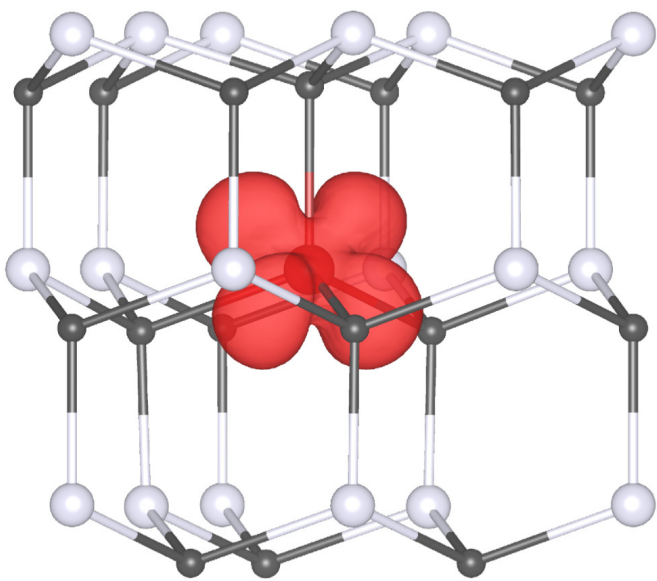 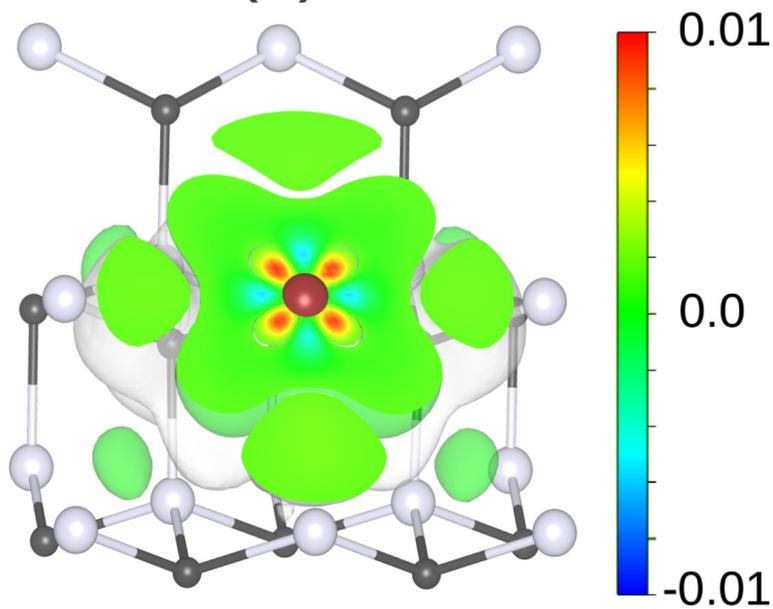



\end{document}